\documentclass[a4paper,11pt]{article}

\usepackage[utf8x]{inputenc}
\usepackage[T1]{fontenc}
\usepackage{color}
\usepackage{amssymb}
\usepackage{amsmath}
\usepackage{graphicx}
\usepackage{mathtools}
\usepackage{ragged2e}

\usepackage{dcolumn}% Align table columns on decimal point
\usepackage{bm}
\usepackage{mathrsfs}
\usepackage{graphicx}
\usepackage{braket}
\usepackage{verbatim} 
\usepackage[english]{babel}
\usepackage{hyperref}
\hypersetup{
citecolor=red,
colorlinks=true,
filecolor=red,
linkcolor=blue,
linktocpage=true,
urlcolor=blue
} 
\usepackage[titletoc,toc,title]{appendix}
\numberwithin{equation}{section}
\usepackage{cleveref}
\usepackage[margin=2.5cm]{geometry}
\usepackage{cite}
\usepackage{epsfig}
\usepackage{float}
\usepackage{cancel}
\usepackage{amsfonts}
\usepackage{enumitem}
\usepackage[font={footnotesize,it}]{caption}
\usepackage{authblk}
{\makeatletter \g@addto@macro\bfseries{\boldmath} \makeatother}

\begin{document}

\title{ Covariant holographic negativity from the entanglement wedge in AdS$_3/$CFT$_2$ }

\author[]{Jaydeep Kumar Basak\thanks{\noindent E-mail:~ {\tt jaydeep@iitk.ac.in}}}
\author[]{Himanshu Parihar\thanks{\noindent E-mail:~ {\tt himansp@iitk.ac.in}}}
\author[]{Boudhayan Paul\thanks{\noindent E-mail:~ {\tt paul@iitk.ac.in}} }
\author[]{Gautam Sengupta\thanks{\noindent E-mail:~ {\tt sengupta@iitk.ac.in}}}

\affil[]{
Department of Physics\\

Indian Institute of Technology\\ 

Kanpur, 208 016, India
}

\date{}

\maketitle

\thispagestyle{empty}

\begin{abstract}

\noindent

\justify

We propose a covariant holographic construction based on the extremal entanglement wedge cross section, for the entanglement negativity of bipartite states in $CFT_{1+1}$s dual to non static bulk $AdS_3$ geometries. Utilizing our proposal we obtain the holographic entanglement negativity for bipartite mixed states in $CFT_{1+1}$s dual to bulk extremal and non extremal rotating BTZ black holes and the time dependent Vaidya-$AdS_3$ geometries. Our results correctly reproduce the replica technique results modulo certain constants for the corresponding dual $CFT_{1+1}$s in the large central charge limit for the bulk black hole geometries. Furthermore for all the cases our results also match upto such constants with the corresponding results from an earlier alternate covariant holographic construction in the literature. This demonstrates the equivalence modulo constants of the two constructions for the $AdS_3/CFT_2$ scenario.

\end{abstract}

\clearpage

\tableofcontents

\clearpage

\section{Introduction}
\label{sec1}

\justify

Characterization of mixed state entanglement for bipartite states in extended quantum many body systems has garnered extensive interest in the investigation of diverse physical phenomena which extends from condensed matter physics to quantum gravity and black holes. This is a subtle and complex issue in quantum information theory as the standard measure  of the entanglement entropy valid for pure states, receives contributions from irrelevant correlations for mixed states. In this context the entanglement (logarithmic) negativity was introduced as a computable measure for mixed state entanglement in a significant work by Vidal and Werner \cite{PhysRevA.65.032314}
which provided an upper bound on the distillable entanglement \footnote {Note that there are other proposed measures to characterize mixed state entanglement in quantum information theory. However most of these involve optimization over LOCC protocols and are not easily computable.}. It was subsequently demonstrated by Plenio \cite{PhysRevLett.95.090503} that although the above measure was non convex it was an entanglement monotone under local operations and classical communication (LOCC) and satisfied the additivity property.

Remarkably it was possible to compute the entanglement negativity for bipartite states in $(1+1)$-dimensional conformal field theories ($CFT_{1+1}$) through a replica technique \cite {Calabrese:2012ew,Calabrese:2012nk,Calabrese:2014yza} modified from a similar technique for the entanglement entropy \cite {Calabrese:2004eu,Calabrese:2009ez,Calabrese:2009qy,Calabrese:2010he}. In this context in a classic work Ryu and Takayanagi (RT) \cite{Ryu:2006bv,Ryu:2006ef} proposed a holographic entanglement entropy conjecture for bipartite states in 
$CFT_d$s dual to bulk $AdS_{d+1}$ geometries through the $AdS/CFT$ correspondence, which involved the area of a 
co-dimension two bulk static minimal surface homologous to the subsystem in question. Subsequently the RT conjecture was utilized to compute the entanglement entropy for various bipartite states in holographic $CFTs$ \cite{Takayanagi:2012kg,Nishioka:2009un,e12112244,Blanco:2013joa,Fischler:2012ca,Fischler:2012uv,Chaturvedi:2016kbk, Nishioka:2018khk} (and references therein). The RT conjecture was generalized to non static bulk $AdS$ geometries in an important work by Hubeny, Rangamani and Takayanagi (HRT) \cite{Hubeny:2007xt} which involved the area of a co dimension two bulk extremal surface homologous to the subsystem under consideration. The HRT conjecture was subsequently employed to compute the covariant entanglement entropy of various bipartite states including time dependent states in holographic $CFT$s. The RT  conjecture was proved for the $AdS_3/CFT_2$ scenario in \cite {Faulkner:2013yia} and in \cite {Lewkowycz:2013nqa,Dong:2016hjy} for a generic $AdS_{d+1}/CFT_d$ framework. Following this a proof for the HRT conjecture was established in \cite {Dong:2016hjy}.

Given the exciting progress described above in the holographic characterization of entanglement entropy in dual $CFT_d$s it was a critical issue to obtain a corresponding holographic description for the entanglement negativity. Following
preliminary investigations in this direction reported in \cite {Rangamani:2014ywa} a proposal for the holographic entanglement negativity was advanced for various bipartite states in a $CFT_{1+1}$ \cite{Chaturvedi:2016rcn,Malvimat:2017yaj,Jain:2017sct,Malvimat:2018txq} in the $AdS_3/CFT_2$ scenario. This involved specific algebraic sums of the lengths of bulk geodesics homologous to appropriate combinations of the intervals in question and exactly reproduced the corresponding $CFT_{1+1}$ replica technique results in the large central charge limit. Subsequently a similar proposal for the covariant holographic entanglement negativity for dual bulk non static $AdS$ geometries was proposed in \cite{Chaturvedi:2016opa,Jain:2017uhe,Malvimat:2018ood} which also reproduced 
the corresponding $CFT_{1+1}$ replica technique results in the large central charge limit for various bipartite states including time dependent states in $CFT_{1+1}$ dual to bulk Vaidya-$AdS_3$ geometries. A higher dimensional extension of the above construction was proposed in \cite{Chaturvedi:2016rft,Parul,Jain:2018bai,Basak:2020bot} for a generic $AdS_{d+1}/CFT_d$ scenario. A bulk proof of the above holographic conjecture, based on replica symmetry breaking saddles for the bulk gravitational path integral in \cite{Dong:2021clv} has been proposed recently in \cite{KumarBasak:2020ams} for the $AdS_{d+1}/CFT_d$ scenario. Note that this proof is restricted to spherically entangling surfaces whereas it remains a non trivial open issue for generic subsystem geometries.

In a related development the concept of an entanglement wedge as a bulk subregion dual to the reduced density matrix in the boundary $CFT$ was developed in \cite{Czech:2012bh,Wall:2012uf,Headrick:2014cta,Jafferis:2014lza,Jafferis:2015del,Dong:2016eik,Faulkner:2017vdd,Harlow:2018fse}. The minimal entanglement wedge cross section (EWCS) has been proposed to be the holographic dual of the entanglement of purification (EoP) \cite{Takayanagi:2017knl,Nguyen:2017yqw} (refer to \cite{Bao:2017nhh,Bhattacharyya:2018sbw,Hirai:2018jwy,Espindola:2018ozt,Umemoto:2018jpc,Bao:2018gck,Agon:2018lwq,BabaeiVelni:2019pkw,Jokela:2019ebz,Guo:2019pfl,Bao:2019wcf,Harper:2019lff,Umemoto:2019jlz,BabaeiVelni:2020wfl,Khoeini-Moghaddam:2020ymm} for recent progress in this direction). Further the connection of the minimal entanglement wedge cross section to the {\it odd entanglement entropy} \cite{Tamaoka:2018ned} and the {\it reflected entropy} \cite{Dutta:2019gen,Engelhardt:2017aux,Engelhardt:2018kcs,Jeong:2019xdr,Bao:2019zqc,Chu:2019etd} has also been established.
%The entanglement of purification (EoP) is described in quantum information theory as a measure of both quantum and classical correlations in a mixed state\cite{Terhal_2002}. 

Following the developments described above, an interesting alternative conjecture for the holographic entanglement negativity for bipartite states in dual $CFT$s involving the minimal entanglement wedge cross section (EWCS) backreacted by the bulk cosmic brane for the conical defect, was proposed in \cite{Kudler-Flam:2018qjo} for subsystems with spherical entangling surfaces in the $AdS/CFT$ framework. Subsequently a ``proof'' of the above conjecture involving the {\it reflected entropy} \cite {Dutta:2019gen} was established in \cite{Kusuki:2019zsp}\footnote {\label {Hayden} Note that the holographic conjecture involving the minimal EWCS described in  \cite{Kudler-Flam:2018qjo,Kusuki:2019zsp} and its proof based on the reflected entropy should also include the Markov gap discussed in a recent communication \cite{Hayden:2021gno}.}. Utilizing their conjecture the authors in \cite{Kudler-Flam:2018qjo,Kusuki:2019zsp} obtained the holographic entanglement negativity for bipartite states in $CFT_{1+1}$s dual to
static bulk $AdS_3$ geometries which matched with the corresponding field theory replica technique results described in \cite{Calabrese:2012ew,Calabrese:2012nk,Calabrese:2014yza} modulo certain constants \footnote { \label{Markov} These constant terms appear to be related to the holographic Markov gap described in \cite{Hayden:2021gno} and mentioned in  footnote \ref {Hayden}.}. Their construction was further refined in \cite {Basak:2020oaf} to resolve an issue for the entanglement negativity of the mixed state configuration of a single interval at a finite temperature in the dual $CFT_{1+1}$ through the appropriate correct construction for the corresponding minimal bulk EWCS. Interestingly these results also matched upto the constants mentioned above, with those obtained through the earlier holographic proposal involving the lengths of bulk geodesics homologous to combinations of intervals described in \cite {Chaturvedi:2016rcn,Malvimat:2017yaj,Jain:2017sct,Malvimat:2018txq}.

Note that the proposal for the holographic entanglement negativity from the EWCS described in \cite{Kudler-Flam:2018qjo, Kusuki:2019zsp} involved bipartite states in $CFT_{1+1}$ dual to bulk static $AdS_3$ geometries. This naturally poses the interesting question of a covariant extension of their construction for $CFT_{1+1}$s dual to non static and time dependent bulk $AdS_3$ space time. In this article we address this significant issue and propose a covariant construction for the holographic entanglement negativity of bipartite mixed states in $CFT_{1+1}$  dual to bulk non static and time dependent $AdS_3$ geometries from the extremal EWCS in an $AdS_3/CFT_2$ scenario. In this context we compute the extremal EWCS for mixed state configurations in $CFT_{1+1}$s dual to bulk non extremal and extremal rotating BTZ black holes employing the techniques described in \cite {Hubeny:2007xt}. From the bulk extremal EWCS obtained through the above construction we subsequently compute the covariant holographic entanglement negativity for the mixed state configurations in question. Remarkably our results for the holographic entanglement negativity from the bulk extremal EWCS reproduces the corresponding field theory replica technique results modulo constants, for these mixed states in the dual $CFT_{1+1}$ in the large central charge limit. Furthermore as earlier for the static case, they also match with the results upto these constants from the earlier alternate approach based on the algebraic sum of the lengths of bulk geodesics homologous to combinations of appropriate intervals described in \cite{Chaturvedi:2016opa,Jain:2017uhe,Malvimat:2018ood}. Naturally this constitutes a strong substantiation of our covariant holographic proposal.

Subsequent to the above analysis for stationary bulk configurations we turn our attention to utilize our covariant construction for the holographic entanglement negativity for bipartite mixed states in $CFT_{1+1}$s dual to time dependent bulk $AdS_3$ geometries. In this context we obtain the time dependent holographic entanglement negativity utilizing our covariant construction based on the bulk extremal EWCS, for bipartite mixed states in $CFT_{1+1}$s dual to  bulk Vaidya-$AdS_3$ geometries through the techniques of \cite {Hubeny:2007xt}. Once again our results match modulo the constants mentioned above with those obtained from the alternate covariant holographic entanglement negativity conjecture mentioned earlier and described in \cite {Chaturvedi:2016opa,Jain:2017uhe,Malvimat:2018ood} which constitutes further consistency checks and a robust substantiation for our covariant construction.
 
This article is organized as follows. In section \ref{sec2} we briefly review the entanglement negativity in quantum information theory. Following this in section \ref{sec3} we recapitulate the computation for the bulk minimal EWCS
and the holographic entanglement negativity conjecture for the static case. In section \ref{sec4} we present our proposal for the covariant holographic entanglement negativity construction based on the extremal bulk EWCS. In section \ref{sec5} we utilize our construction to obtain the holographic entanglement negativity for bipartite mixed states in $CFT_{1+1}$s dual to bulk non extremal and extremal rotating BTZ black holes. Subsequently in section \ref{sec6} we employ our proposal to obtain the covariant holographic entanglement negativity for bipartite mixed states in $CFT_{1+1}$s dual to bulk time dependent Vaidya-$AdS_3$ geometries. Finally in section \ref{sec7} we present our summary and discussions.

\section{Entanglement negativity}\label{sec2}

In this section we briefly review the definition of entanglement negativity in quantum information theory \cite{Vidal}(see \cite{Rangamani:2014ywa} for a review). In this context a tripartite system in a pure state is considered which is constituted by the subsystems $A_1$, $A_2$ and $B$, where $A=A_1\cup A_2$ and $B=A^c$ represent the rest of the system. For the Hilbert space $\mathcal{H}=\mathcal{H}_1 \otimes \mathcal{H}_2$, the reduced density matrix for the subsystem $A$ which is in a mixed state is defined as $\rho_{A}=\mathrm{Tr}_{B} \,\rho $   and  $\rho_{A}^{T_2}$ is the partial transpose of the reduced density matrix with respect to the subsystem $A_2$ which is given as
\begin{equation}
\langle e^{(1)}_ie^{(2)}_j|\rho_A^{T_2}|e^{(1)}_ke^{(2)}_l\rangle = 
\langle e^{(1)}_ie^{(2)}_l|\rho_A|e^{(1)}_ke^{(2)}_j\rangle,
\end{equation}
where $|e^{(1)}_i\rangle$ and $|e^{(2)}_j\rangle$ are the bases for the Hilbert spaces
$\mathcal{H}_1$ and  $\mathcal{H}_2$. The entanglement negativity for the bipartite mixed state configuration $A\equiv A_1 \cup A_2$ may then be defined as the logarithm of the trace norm of the partially transposed reduced density matrix  as
\begin{equation}
\mathcal{E} = \ln \mathrm{Tr}|\rho_A^{T_2}|,
\end{equation}
where the trace norm $\mathrm{Tr}|\rho_A^{T_2}|$ is given by the sum of absolute eigenvalues of $\rho_A^{T_2}$. The entanglement negativity satisfies certain properties in quantum information theory namely it is additive, monotonic, and obeys the monogamy inequality however it is not convex. It is a computable measure which provides the upper bound on the distillable entanglement of the mixed state under consideration. Interestingly for bipartite states in $(1+1)$-dimensional conformal field theories($CFT_{1+1}$s) it is possible to explicitly compute the entanglement negativity through a replica technique as described in \cite{Calabrese:2012ew,Calabrese:2012nk,Calabrese:2014yza}. This technique involves the quantity $\mathrm{Tr} \big(\,\rho_A^{T_{2}}\big)^{n}$ for even $n=n_e$ and its analytic continuation to $n_e\to 1$ which leads to the following expression
\begin{equation}
\mathcal{E} = \lim_{n_e \rightarrow 1}  \ln \Big[ \mathrm{Tr} \big(\,\rho_A^{T_{2}}\big)^{n_e} \Big].
\end{equation}

\section{Review of entanglement wedge }\label{sec3}
In this section we briefly review the evaluation of the bulk entanglement wedge cross section \cite {Czech:2012bh,Wall:2012uf,Headrick:2014cta,Jafferis:2014lza,Jafferis:2015del,Takayanagi:2017knl,Nguyen:2017yqw,Bhattacharyya:2018sbw,Bao:2017nhh,Hirai:2018jwy,Espindola:2018ozt,Umemoto:2018jpc,Bao:2018gck,Umemoto:2019jlz,Guo:2019pfl,Bao:2019wcf,Harper:2019lff,Terhal_2002,Tamaoka:2018ned,Dutta:2019gen,Jeong:2019xdr,Bao:2019zqc,Chu:2019etd} for the $AdS_{d+1}/CFT_d$ scenario. Subsequently we
review a conjecture for the holographic entanglement negativity of bipartite states in a $CFT_{1+1}$ related to the above construction in the $AdS_3/CFT_2$ scenario for static cases described in \cite {Kudler-Flam:2018qjo,Kusuki:2019zsp,Basak:2020oaf}.

%Subsequently we
%review a conjecture for the holographic entanglement negativity of bipartite states in a $CFT_{1+1}$ related to the above construction in the $AdS_3/CFT_2$ scenario for static cases described in \cite {Kudler-Flam:2018qjo,Kusuki:2019zsp,Basak:2020oaf}.

\subsection{Entanglement wedge}
\begin{figure}[H]
\centering
\includegraphics[scale=.45 ]{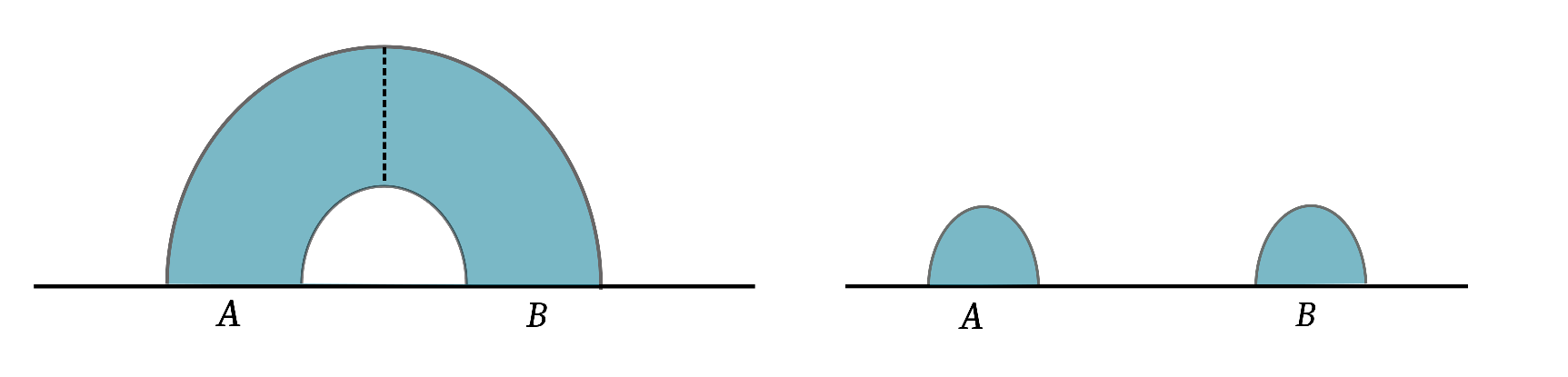}
\caption{Left: The coloured region represents the entanglement wedge for subsystem $A\cup B$ in Poincar\'e $AdS_3$. The dotted line shows the minimal entanglement wedge cross section. Right: If $A$ and $B$ are sufficiently far away, the entanglement wedge becomes disconnected and $E_W(A:B)=0$.}
\label{fig1}
\end{figure}

Consider the spatial subsystems $A$ and $B$ in a $CFT_d$ dual to a static bulk $AdS_{d+1}$ geometry. Let $\Xi$ be a constant time slice in the bulk and $\Gamma_{AB}$ be the RT surface for the subsystem $A\cup B$. Then the co dimension 1 spatial region in $\Xi$ bounded by $A\cup B\cup\Gamma_{AB}$ is defined as the entanglement wedge in the bulk geometry. The minimal EWCS is the minimal area surface $\Sigma_{AB}^{min}$ which splits the entanglement wedge $\Xi$ in two parts containing $A$ and $B$ separately. This is illustrated in Fig. \ref{fig1}. 
It is important to note here that the reduced density matrix $\rho_{AB}$ has been shown to be dual to the corresponding entanglement wedge \cite{Czech:2012bh,Wall:2012uf,Headrick:2014cta,Jafferis:2014lza,Jafferis:2015del,Dong:2016eik,Faulkner:2017vdd,Harlow:2018fse}. Now for the above case
the minimal entanglement wedge cross section $E_W$ (min EWCS) is defined as follows
\begin{equation}\label{static-wedge}
E_W(A:B)=\frac{\mathrm{Area}(\Sigma_{AB}^{min})}{4 G_N},
\end{equation}
where $G_N$ is Newton's constant.  
Note that the entanglement wedge describes the bulk region defined as the domain of dependence of the homology surface $R_A$ which is a space-like surface  bounded by the subsystem $A$ and the RT surface $\Gamma_A$. For static  bulk geometries the entanglement wedge is defined on a constant time slice. In this case the authors in \cite{Takayanagi:2017knl,Nguyen:2017yqw} have conjectured that the minimal entanglement wedge cross section (min EWCS) is dual to the {\it entanglement of purification} (EoP) in quantum information theory as it satisfies identical properties.

We now proceed to briefly review the computation of the minimal entanglement wedge cross section for various configurations in the $AdS_3/CFT_2$ scenario.
In this context, we consider the zero temperature mixed state configuration described by the subsystems $A$ and $B$ for the $CFT_{1+1}$ vacuum which are characterized by disjoint spatial intervals $A=[a_1,a_2]$ and $B=[b_1,b_2]$ having lengths $l_1$ and $l_2$ respectively separated by a distance $l_s$. The bulk dual for this case is a pure $AdS_3$ space time in Poincar\'e coordinates whose metric (with $AdS$ radius $R=1$) at a constant time slice given by the metric
\begin{equation}
ds^2=\frac{dx^2+dz^2}{z^2}.
\end{equation}   
Note that when the mutual information ${\cal I}(A:B)\equiv [S(A)+S(B)-S(A\cup B) ]>0$, the entanglement wedge remains connected otherwise it is disconnected. The minimal entanglement wedge cross section\footnote{Note that it can be written in a different form \cite{Kudler-Flam:2018qjo,Kudler-Flam:2019wtv}.
\begin{equation}\label{ryu-ew}
E_W=
\left\{
\begin{array}{ll}
\displaystyle
\frac{c}{6} \ln \frac{1+\sqrt{x}}{1-\sqrt{x}},
&
1/2 \leq x \leq 1,
\\
\\
0,
&
0 \leq x \leq 1/2,
\end{array}
\right.
\end{equation}
where $x=\frac{(a_2-a_1)(b_2-b_1)}{(b_1-a_1)(b_2-a_2)}=\frac{l_1 l_2}{(l_1+l_s)(l_2+l_s)}$ is the cross ratio related to $y$ by $y=\frac{x}{1-x}$.} for this configuration after using the Brown-Henneaux formula \cite{Brown:1986nw} may be expressed as \cite{Takayanagi:2017knl}
\begin{equation}\label{min-cross-section}
E_W(A:B)=
\left\{
\begin{array}{ll}
\displaystyle
\frac{c}{6}\ln\left(1+2y+2\sqrt{y(y+1)}\right), 
&
y>1,
\\ \\
0,
&
0\leq y\leq 1.
\end{array}
\right.
\end{equation}
Here $y$ is the cross ratio given by
\begin{equation}
y=\frac{(a_2-a_1)(b_2-b_1)}{(b_1-a_2)(b_2-a_1)}=\frac{l_1 l_2}{l_s \,(l_1+l_2+l_s)}.
\end{equation}

For the corresponding finite temperature mixed state configuration in a  $CFT_{1+1}$ defined on an infinite line, the bulk dual is a planar BTZ black hole (black string) whose metric is given as
\begin{equation}\label{static-btz}
ds^2=\frac{1}{z^2}\left(-f(z)dt^2+\frac{dz^2}{f(z)}+dx^2\right),
\,\, f(z)\equiv1-z^2/z_H^2,
\end{equation}
where the event horizon is located at $z=z_H$ and the inverse temperature $\beta$ may be expressed as $\beta=2\pi z_H$. The minimal entanglement wedge cross section corresponding to the disjoint intervals $A=[a_1,a_2]$ and $B=[b_1,b_2]$ in a dual $CFT_{1+1}$ is given as follows \cite{Takayanagi:2017knl}
\begin{equation}\label{cross-section-temp}
E_W(A:B)= \frac{c}{6}
\ln\left(1+2\zeta+2\sqrt{\zeta(\zeta+1)}\right),
\end{equation}
with $\zeta$ being given as
\begin{equation}
\zeta\equiv\frac{\sinh\left(\frac{\pi(a_{2}-a_{1})}{\beta}\right)\sinh\left(\frac{\pi(b_{2}-b_{1})}{\beta}\right)}{\sinh\left(\frac{\pi(b_{1}-a_{2})}{\beta}\right)\sinh\left(\frac{\pi(b_{2}-a_{1})}{\beta}\right)}
=\frac{\sinh\left(\frac{\pi l_1}{\beta}\right)\sinh\left(\frac{\pi l_2}{\beta}\right)}{\sinh\left(\frac{\pi l_s}{\beta}\right)\sinh\left(\frac{\pi(l_1+l_2+l_s)}{\beta}\right)}.
\end{equation}

The configuration of adjacent intervals can be obtained by taking the adjacent limit $l_s\to \epsilon$ where $\epsilon$ is a cut off. The minimal entanglement wedge cross section for two adjacent intervals in a pure $AdS_3$ space time may then be obtained from eq. (\ref{min-cross-section}) as
\cite{Kudler-Flam:2018qjo}
\begin{equation}\label{ew-adj}
E_W(A:B)=\frac{c}{6}\ln \left(\frac{l_1 l_2}{\epsilon(l_1+l_2)}\right)+\frac{c}{6}\ln 4,
\end{equation} 
where $\epsilon$ is the UV cut off.
Similarly the minimal entanglement wedge cross section for the adjacent intervals at a finite temperature may be computed from eq. (\ref{cross-section-temp}) by taking the adjacent limit $l_s\to \epsilon$ which may be expressed as \cite{Kudler-Flam:2018qjo}
\begin{equation}\label{ew-adj-finite}
E_W(A:B)=\frac{c}{6}\ln \left[\left(\frac{\beta}{\pi \epsilon}\right)\frac{\sinh\left(\frac{\pi l_1}{\beta}\right)\sinh\left(\frac{\pi l_2}{\beta}\right)}{\sinh\left(\frac{\pi (l_1+l_2)}{\beta}\right)}\right]+\frac{c}{6}\ln 4,
\end{equation}

For the configuration of two disjoint intervals, we only discuss the case where the disjoint intervals are in proximity $(x \to 1)$. Taking the limit $x \to 1$ in eq. (\ref{ryu-ew}), we arrive at the minimal entanglement wedge cross section for the disjoint intervals in proximity as follows
\begin{equation}\label{ew-disj-zero}
E_W(A:B)=\frac{c}{6}\ln\left(\frac{1}{1-x}\right)+\frac{c}{6}\ln 4.
\end{equation}

Finally, the minimal entanglement wedge cross section for a single interval at zero temperature may be obtained from the property of EWCS for a pure state i.e $E_W(A,B)=S(A)$.
The minimal EWCS for a single interval at a finite temperature is computed by considering a tripartite system consisting of an interval $A$ of length $l$ with two auxiliary interval $B_1, B_2$ of length $L$ adjacent on either side of $A$ as described in \cite {Calabrese:2014yza}. Upon using the polygamy property of the minimal EWCS and subsequently implementing the bipartite limit $L\to \infty$, the minimal EWCS for this configuration is given by \cite{Basak:2020oaf}
\begin{equation}\label{ew-BTZ-single}
E_W(A:B)=\frac{c}{3}\ln \left(\frac{\beta}{\pi\epsilon} \sinh\frac{\pi l}{\beta}\right)-\frac{\pi c l}{3 \beta}+\frac{c}{3}\ln 4.
\end{equation}

\subsection{Holographic entanglement negativity from the EWCS}
For the $CFT$s dual to bulk static $AdS$ configurations, the authors in \cite{Kudler-Flam:2018qjo,Kusuki:2019zsp} motivated by results from  quantum error-correcting codes, proposed that the holographic entanglement negativity for $CFT_d$s dual to bulk static $AdS_{d+1}$ geometries, could be obtained from the area of the backreacted minimal entanglement wedge cross section (EWCS) for spherical entangling surfaces. Note that the 
EWCS was also shown to be related to the holographic entanglement of purification (EoP)\cite{Takayanagi:2017knl,Nguyen:2017yqw}. For spherical entangling surfaces the holographic entanglement negativity for the dual $CFT_d$s could be expressed in terms of the bulk minimal entanglement wedge cross section backreacted by the cosmic brane arising from the conical defect as follows 
\cite{Kudler-Flam:2018qjo,Kusuki:2019zsp} 
\begin{equation}\label{holo-neg}
\mathcal{E}= \mathcal{X}_d E_W.
\end{equation}
Here $\mathcal{X}_d$ is a constant which arises from the backreaction of the bulk cosmic brane for the conical defect in the replica limit ($n\to \frac{1}{2}$)  
\cite {Dong:2016fnf} and is dependent on the dimension of the bulk space time. For the 
pure vacuum state in a $CFT_d$, this constant could be determined in \cite{Hung:2011nu,Rangamani:2014ywa} as
\begin{equation}
\mathcal{X}_d = \left(\frac{1}{2}x_d^{d-2}\left(1 + x_d^2\right) - 1\right), \,\,\,\,\, x_d = \frac{2}{d}\left(1 + \sqrt{1 - \frac{d}{2} + \frac{d^2}{4}}\right).
\end{equation}
In the context of $AdS_3/CFT_2$ scenario for static bulk geometries, this quantity is given as $\mathcal{X}_2=3/2$ which may also be verified through explicit replica technique calculations from the $CFT_{1+1}$ \cite {Calabrese:2012ew,Calabrese:2012nk}. Utilizing the proposal described above in eq. (\ref {holo-neg}) the authors in \cite {Kudler-Flam:2018qjo,Kusuki:2019zsp} obtained the holographic entanglement negativity for bipartite states in $CFT_{1+1}$s dual to static bulk $AdS_3$ geometries upto certain constants possibly related to the Markov gap described in \cite {Hayden:2021gno} (see footnotes \ref{Hayden} and \ref{Markov} in the Introduction). 

\section{Covariant entanglement negativity in $AdS_3/CFT_2$}\label{sec4}

In this section we proceed to establish a covariant holographic entanglement negativity proposal for mixed states in $CFT_{1+1}$s dual to bulk non-static configurations in the $AdS_3/CFT_2$ framework based on the entanglement wedge cross section (EWCS) . Note that it has been described in 
\cite{Takayanagi:2017knl,Nguyen:2017yqw} that the entanglement wedge cross section may also be generalized to non static bulk configurations. To this end we consider the disjoint spatial subsystems $A$ and $B$ in a dual CFT. Let $\Gamma_{AB}^{ext}$ be the extremal (HRT) surface for the subsystem $A\cup B$. 
The entanglement wedge in this case is the bulk region with the spatial boundary $A\cup B\cup \Gamma_{AB}^{ext}$. Then the extremal entanglement wedge cross section may be defined as the area of the minimal extremal surface $\Sigma_{AB}^{ext}$ in the entanglement wedge which separates $A$ from $B$ and also ends on the HRT surface $\Gamma_{AB}^{ext}$. It is defined as follows
\begin{equation}\label{cov-wedge}
E_W^{ext}(A:B)=\frac{\mathrm{Area}\,(\Sigma_{AB}^{ext})}{4 G_N}.
\end{equation}
Note that the above eq. (\ref{cov-wedge}) reduces to the eq. (\ref{static-wedge}) for the static case. The extremal entanglement wedge cross section for the covariant case possess all the properties of that for the corresponding static case \cite{Takayanagi:2017knl,Nguyen:2017yqw}. 

As described in the previous section, the determination of the holographic entanglement negativity  involves the backreacted minimal EWCS due to the cosmic brane for the conical defect \cite {Kudler-Flam:2018qjo,Kusuki:2019zsp}. This motivates us to propose a covariant holographic proposal for the entanglement negativity in the $AdS_3/CFT_2$ scenario as follows
\begin{equation}\label{cov-neg-proposal}
\mathcal{E}=\frac{3}{2}E_W^{ext}.
\end{equation}
Here $E_W^{ext}$ is the corresponding extremal entanglement wedge cross section for the dual non static bulk configuration where we have assumed that the backreaction parameter $\mathcal{X}_2 =\frac {3}{2}$ also in the covariant $AdS_3/CFT_2$ scenario as the extremal curves are described again by geodesics as in the static case \footnote { An argument for the unchanging backreaction factor due to the cosmic brane may  be inferred from the maximin construction described in \cite {Wall:2012uf} which demonstrates an equivalence of the RT and the HRT constructions through the maximization of the areas of minimal surfaces on distinct Cauchy slices. Note that out of all the possible surfaces involved in the maximin procedure there will always exist a stable extremal surface which will be identified as the corresponding HRT surface. }. This is also supported by the replica technique results for the dual holographic $CFT_2$ \cite{Chaturvedi:2016opa,Jain:2017uhe,Malvimat:2018ood}. 

We should mention here however that an 
explicit analytic continuation describing the backreaction of the cosmic brane on the bulk geometry and the extremal
EWCS along the lines of \cite {Dong:2016fnf,Kudler-Flam:2018qjo,Kusuki:2019zsp} is still a non trivial open issue 
for a generic higher dimensional covariant $AdS_{d+1}/CFT_d$ scenario with arbitrary subsystem geometries.  
With this caveat we now proceed to substantiate our proposal for the $AdS_3/CFT_2$ scenario by computing the covariant holographic entanglement negativity from the bulk EWCS for various mixed state configurations in $CFT_{1+1}$s dual to bulk non-extremal and extremal rotating BTZ black holes and also the time dependent Vaidya-$AdS_3$ geometries in the subsequent sections.

\section{Rotating BTZ black hole}\label{sec5}
In this section we utilize our proposal described above to compute the covariant holographic entanglement negativity for bipartite mixed states in $CFT_{1+1}$s  dual to bulk non-extremal and extremal rotating BTZ black holes from the corresponding extremal entanglement wedge cross sections.

\subsection{Non-extremal rotating BTZ black hole}
The bulk non extremal rotating BTZ black hole is a stationary configuration dual to mixed states in a $CFT_{1+1}$ at a finite temperature with a conserved angular momentum.
The metric for the non-extremal rotating BTZ black hole with a planar horizon (black string) and the AdS length $(R=1)$ is given as
\begin{equation}\label{non-ext}
\begin{aligned}
 ds^{2}&=-\frac{(r^2-r_{+}^2)(r^2-r_{-}^2)}{r^2}dt^2 \\
 &+ \frac{r^2 dr^2}{(r^2-r_{+}^2)(r^2-r_{-}^2)}+ r^2(d\phi-\frac{r_{+}r_{-}}{r^2}dt)^2.
 \end{aligned}
\end{equation}
Here the coordinate $\phi$ is non compact and $r_{\pm}$ are the outer and inner horizon radius of the black hole respectively. The mass $M$, angular momentum $J$, Hawking temperature $T_H$, and the angular velocity $\Omega$ are related to the horizon radii $r_{\pm}$ which can be expressed as follows
\begin{eqnarray}\label{MJT-Omega}
 M = r_{+}^2+r_{-}^2,~~~~~~~~ J=2 r_{+}r_{-},\\
  T_H=\frac{1}{\beta}=\frac{r_{+}^2-r_{-}^2}{2\pi r_+},  ~~~~~\Omega=\frac{r_{-}}{r_{+}},~~~~~~~\beta_{\pm}=\beta(1\pm\Omega).
\end{eqnarray}
The corresponding temperatures relevant to the left and the right moving modes in the dual $CFT_{1+1}$ with a conserved angular momentum defined on the cylinder (twisted) are described as \cite {Chaturvedi:2016opa,Jain:2017uhe,Malvimat:2018ood}
\begin{equation}\label{cfttemp1}
T_{\pm}=\frac{1}{\beta_{\pm}}=\frac{r_+\mp r_-}{2\pi}.
\end{equation}
To obtain the extremal entanglement wedge cross section (EWCS) in this bulk geometry, we map the rotating BTZ metric in eq. (\ref{non-ext}) to the Poincar\'{e} patch of the $AdS_3$ space time as follows
\begin{equation}\label{non-to-poin}
\begin{aligned}
 w_{\pm}=&\sqrt{\frac{r^2-r_{+}^2}{r^2-r_{-}^2}} e^{2\pi T_{\pm}u_{\pm}}\equiv X\pm T,\\
  Z=&\sqrt{\frac{r_{+}^2-r_{-}^2}{r^2-r_{-}^2}} e^{\pi T_+ u_+ + \pi T_- u_- },
\end{aligned}
\end{equation}
where $u_{\pm}=\phi \pm t$. The resulting $AdS_3$ metric in the Poincar\'{e} coordinates is then given by
\begin{equation}\label{poincare-coord}
 ds^2=\frac{dw_{+}dw_{-}+dZ^2}{Z^2}\equiv\frac{-dT^2+dX^2+dZ^2}{Z^2}.
\end{equation}
We now turn our attention to compute the extremal EWCS and corresponding covariant holographic negativity for the mixed state configurations of adjacent, disjoint and a single interval in the following subsections. 

\subsubsection{Covariant holographic negativity for adjacent intervals}\label{subs-adj}
For $AdS_3$ in Poincar\'{e} coordinates described in eq. (\ref{poincare-coord}),
the minimal EWCS for two adjacent intervals $A$ and $B$ on the boundary having lengths $\Delta \phi_1$ and $\Delta \phi_2$ respectively may be expressed from eq. (\ref{ew-adj}) as
\begin{equation}\label{ew-local-poincare}
E_W(A:B)=\frac{c}{6}\ln \left(\frac{\Delta \phi_1 \Delta \phi_2}{\epsilon (\Delta \phi_1+\Delta \phi_2)}\right)+\frac{c}{6}\ln 4.
\end{equation}
The infra red cut-offs $r_\infty$ and $\varepsilon_{i}$ ($i=1,2$) for the bulk in the rotating  BTZ and the Poincar\'{e} coordinates corresponding to an interval $A$ at a fixed time $t_0$ and having length $\Delta \phi=|\phi_1-\phi_2|$ are related as \cite {Hubeny:2007xt}
\begin{equation}\label{cut-off}
\varepsilon_{i}= \sqrt{\frac{r_{+}^2-r_{-}^2}{r_{\infty}^2}}e^{(r_{+}\phi_{i}~-~t_0 r_{-})},
\end{equation}
where $\varepsilon_{i}$ represents the cut off at each of the two endpoints of the interval $A$.
The length of the interval $\Delta \phi$ in the rotating BTZ coordinates may then be expressed as \cite {Hubeny:2007xt}
\begin{equation}\label{relation}
(\Delta\phi)^2=\Delta w_+ \Delta w_{-}=(e^{(r_+-r_{-})(\phi_1+t_0)}-e^{(r_+-r_-)(\phi_2+t_0)})
(e^{(r_++r_-)(\phi_1-t_0)}-e^{(r_++r_-)(\phi_2-t_0)}).
\end{equation}
Following the procedure explained in \cite {Hubeny:2007xt}, we obtain the extremal EWCS corresponding to adjacent intervals of lengths $\Delta \phi_1=l_1$ and $\Delta \phi_2=l_2$ in terms of the original rotating BTZ coordinates using eqs. (\ref{cut-off}) and (\ref{relation}) in eq. (\ref{ew-local-poincare}) as
\begin{equation}\label{nonextnegbulk}
\begin{aligned}
E_W^{ext}(A:B)&=\frac{c}{12}\ln\Bigg[\bigg(\frac{\beta_{+}}{\pi \epsilon}\bigg)\frac{\sinh{\big(\frac{\pi l_1}{\beta_{+}}\big)}\sinh{\big(\frac{\pi l_2}{\beta_{+}}\big)}}{\sinh\{{\frac{\pi}{\beta_{+}} (l_1+l_2)}\}}\Bigg]\\
&+\frac{c}{12}\ln\Bigg[\bigg(\frac{\beta_{-}}{\pi \epsilon}\bigg)\frac{\sinh{\big(\frac{\pi l_1}{\beta_{-}}\big)}\sinh{\big(\frac{\pi l_2}{\beta_{-}}\big)}}{\sinh\{{\frac{\pi}{\beta_{-}} (l_1+l_2)}\}}\Bigg]+\frac{c}{6}\ln 4,
\end{aligned}
\end{equation}
where $\epsilon$ is the UV cut-off for the dual $CFT_{1+1}$ which is related to the bulk infra red cut-off $r_\infty$ as $\epsilon={1}/{r_\infty}$.

We now compute the covariant holographic negativity using our construction given in eq. (\ref{cov-neg-proposal})
for the mixed state of adjacent intervals with lengths $l_1$ and $l_2$ in a  $CFT_{1+1}$ with a conserved angular momentum (charge) dual to the bulk non-extremal BTZ black hole as follows
\begin{equation}\label{en-non-ext}
\begin{aligned}
{\cal E}&=\frac{c}{8}\ln\Bigg[\bigg(\frac{\beta_{+}}{\pi \epsilon}\bigg)\frac{\sinh{\big(\frac{\pi l_1}{\beta_{+}}\big)}\sinh{\big(\frac{\pi l_2}{\beta_{+}}\big)}}{\sinh\{{\frac{\pi}{\beta_{+}} (l_1+l_2)}\}}\Bigg]\\
&+\frac{c}{8}\ln\Bigg[\bigg(\frac{\beta_{-}}{\pi \epsilon}\bigg)\frac{\sinh{\big(\frac{\pi l_1}{\beta_{-}}\big)}\sinh{\big(\frac{\pi l_2}{\beta_{-}}\big)}}{\sinh\{{\frac{\pi}{\beta_{-}} (l_1+l_2)}\}}\Bigg]+\frac{c}{4}\ln 4,\\
&=\frac{1}{2}({\cal E}_L+{\cal E}_R).
\end{aligned}
\end{equation}
We observe that the above covariant holographic negativity matches with corresponding $CFT_{1+1}$ replica result \cite{Jain:2017uhe} upto a constant\footnote{See footnote \ref{Markov}.} and  decouples into ${\cal E}_L$ and ${\cal E}_R$  with the left and right moving inverse temperatures $\beta_{+}$ and $\beta_{-}$ respectively \cite{Eisler_2014}.
Note that the above result in eq. (\ref{en-non-ext}) modulo the constant may also be obtained using an alternative proposal \cite{Jain:2017uhe} which involves the combination of bulk extremal curves (geodesics) homologous to appropriate combinations of the intervals. This is given as
\begin{equation}\label{CO-HEN-CONJ}
\begin{aligned}
{\cal E}&= \frac{3}{16G_{N}^{(3)}} \Big[{\cal L}_{A^t}^{ext}+{\cal L}_{B^t}^{ext}-{\cal L}^{ext}_{A^t\cup B^t}\Big],
\end{aligned}
\end{equation}
where ${\cal L}_{\gamma}^{ext}$ ($\gamma \in A^t, B^t, A^t\cup B^t $) are the lengths of bulk extremal curves homologous to the respective intervals and their combinations.

\subsubsection{Covariant holographic negativity for disjoint intervals}\label{subs-disj}
To obtain the extremal EWCS for disjoint intervals in proximity in the $CFT_{1+1}$ dual to the bulk non-extremal rotating BTZ black hole, we use the Poincar\'{e} coordinates for the bulk $AdS_3$ geometry as described in eq. (\ref{poincare-coord}). To this end we consider the configuration of two disjoint intervals $A$ and $B$ having lengths $l_1$, $l_2$ respectively which are separated by the interval $A_s$ with length $l_s$. Utilizing the eqs. (\ref{ew-disj-zero}), (\ref{cut-off}) and (\ref{relation}), the extremal EWCS in this bulk geometry for configuration of two disjoint intervals in proximity may written as

\begin{equation}
\begin{aligned}
E_W^{ext}(A:B)
&= \frac{ c }{12} \ln \left [ \frac
{
\sinh \left \{ \frac{ \pi \left ( l_1 + l_s \right ) }{ \beta_{+} } \right \}
\sinh \left \{ \frac{ \pi \left ( l_2 + l_s \right ) }{ \beta_{+} } \right \}
}{
\sinh \left ( \frac{ \pi l_s }{ \beta_{+} } \right )
\sinh \left \{ \frac{ \pi \left ( l_1 + l_2 + l_s \right ) }{ \beta_{+} } \right \}
} \right ] \\
&  + \frac{ c }{12 } \ln \left [ \frac
{
\sinh \left \{ \frac{ \pi \left ( l_1 + l_s \right ) }{ \beta_{-} } \right \}
\sinh \left \{ \frac{ \pi \left ( l_2 + l_s \right ) }{ \beta_{-} } \right \}
}{
\sinh \left ( \frac{ \pi l_s }{ \beta_{-} } \right )
\sinh \left \{ \frac{ \pi \left ( l_1 + l_2 + l_s \right ) }{ \beta_{-} } \right \}
} \right ] +\frac{c}{6}\ln 4\\
\end{aligned}
\end{equation}
We now use the proposal described in eq. (\ref{cov-neg-proposal}) to determine the covariant holographic negativity for the mixed state of two disjoint interval (in proximity) in the dual $CFT_{1+1}$ as follows
\begin{equation}\label{neg-non-ext-disj}
\begin{aligned}
{\cal E}
& = \frac{ c }{ 8 } \ln \left [ \frac
{
\sinh \left \{ \frac{ \pi \left ( l_1 + l_s \right ) }{ \beta_{+} } \right \}
\sinh \left \{ \frac{ \pi \left ( l_2 + l_s \right ) }{ \beta_{+} } \right \}
}{
\sinh \left ( \frac{ \pi l_s }{ \beta_{+} } \right )
\sinh \left \{ \frac{ \pi \left ( l_1 + l_2 + l_s \right ) }{ \beta_{+} } \right \}
} \right ] \\
& + \frac{ c }{ 8 } \ln \left [ \frac
{
\sinh \left \{ \frac{ \pi \left ( l_1 + l_s \right ) }{ \beta_{-} } \right \}
\sinh \left \{ \frac{ \pi \left ( l_2 + l_s \right ) }{ \beta_{-} } \right \}
}{
\sinh \left ( \frac{ \pi l_s }{ \beta_{-} } \right )
\sinh \left \{ \frac{ \pi \left ( l_1 + l_2 + l_s \right ) }{ \beta_{-} } \right \}
} \right ] +\frac{c}{4}\ln 4\\
& = \frac{ 1 }{ 2 } \left ( {\cal E}_L + {\cal E}_R \right ).
\end{aligned}
\end{equation}
As earlier it is observed that the above result matches with the corresponding $CFT_{1+1}$ replica result \cite{Malvimat:2018ood} upto a constant and is independent of the UV cut of{}f $\epsilon$.
Once more the covariant holographic entanglement negativity decouples into $ {\cal E}_L $ and $ {\cal E}_R $ with the left and right moving inverse temperatures $ \beta_{+} $ and $ \beta_{-} $ respectively \cite{Eisler_2014}. 
Note that the covariant holographic entanglement negativity for the two disjoint intervals in proximity from eq. (\ref{neg-non-ext-disj}) may also be obtained modulo the constant from an earlier holographic proposal in \cite{Malvimat:2018ood} which involves the combination of bulk extremal curve anchored on respective intervals and their combinations. This proposal is described as follows
\begin{equation}\label{cohendisj}
\begin{aligned}
{\cal E} &= \frac{ 3 } { 16 G_N^{(3)} }
\left ( {\cal L}_{ A^t \cup A_s^t }^{ext}
+ {\cal L}_{ B^t \cup A_s^t }^{ext}
- {\cal L}_{ A^t \cup B^t \cup A_s^t }^{ext}
- {\cal L}_{ A_s^t }^{ext}
\right ),
\end{aligned}
\end{equation}
where $ {\cal L}_{\gamma}^{ext} $ represents the length of the bulk extremal curves homologous to the intervals and their combinations in the dual $CFT_{1+1}$.

\subsubsection{Covariant holographic negativity for a single interval}\label{subs-single}
For this case we start with the $AdS_3$ in Poincar\'{e} coordinates as given in eq. (\ref{poincare-coord}) and consider a tripartite system consisting of an interval $A$ of length $l$ and two adjacent auxiliary intervals of length $L$ on either side of $A$. The minimal EWCS in the Poincar\'{e} coordinates for a single interval may then be expressed using the polygamy property of EWCS and eq. (\ref{ew-adj}) as
\begin{equation}\label{adj1}
E_W(A:B_1B_2)=\frac{c}{3}\ln \left(\frac{l~ L}{\epsilon (L+l)}\right) +\frac{c}{3}\ln 4.
\end{equation}
On using eqs. (\ref{cut-off}) and (\ref{relation}) in the eq. (\ref{adj1}) above followed by implementing the bipartite limit $L\to\infty$, we obtain the extremal EWCS in the original BTZ coordinates as
\begin{equation}
E_W^{ext}(A:B)= \frac{c}{6}\ln\left[\frac{\beta_{+}\beta_{-}}{\pi^2\epsilon^2}\sinh{\left(\frac{\pi l}{\beta_{+}}\right)}\sinh{\left(\frac{\pi l}{\beta_{-}}\right)}\right]-\frac{\pi c l}{3\beta(1-\Omega^2)}+\frac{c}{3}\ln 4.
\end{equation}
We now utilize the holographic proposal eq. (\ref{cov-neg-proposal}) to evaluate the covariant holographic entanglement negativity of the single interval $A$ in the dual $\mathrm{\mathrm{\mathrm{CFT_{1+1}}}}$ as
\begin{equation}\label{en-non-ext-single}
\mathcal{E}= \frac{c}{4}\ln\left[\frac{\beta_{+}\beta_{-}}{\pi^2\epsilon^2}\sinh{\left(\frac{\pi l}{\beta_{+}}\right)}\sinh{\left(\frac{\pi l}{\beta_{-}}\right)}\right]-\frac{\pi c l}{2\beta(1-\Omega^2)}+\frac{c}{2}\ln 4.
\end{equation}
The above result matches with the corresponding $CFT_{1+1}$ replica result \cite{Chaturvedi:2016opa} upto a constant as earlier (See footnote \ref{Markov}).
Note that the expression for the covariant holographic entanglement negativity above may be expressed in terms of the entanglement entropy  and the thermal entropy of the interval $A$ as
\begin{equation}
 {\cal E}= \frac{3}{2}\bigg[S_A-S^{th}_A\bigg]+const.
\end{equation}
This clearly demonstrates that the covariant holographic entanglement negativity captures the distillable entanglement through the subtraction of the contribution from the thermal correlations. Note that this is an universal result for this finite temperature mixed state configuration as described in \cite {Chaturvedi:2016opa} where we have neglected the non universal contributions which are sub leading in $\frac {1}{c}$.

It must be mentioned here that the covariant entanglement negativity for a single interval described in eq. (\ref{en-non-ext-single}) above may also be obtained upto the constant factor from an earlier alternative holographic construction involving the algebraic sum of the lengths of bulk extremal curves homologous to appropriate intervals and their combinations as follows \cite{Chaturvedi:2016opa}
\begin{equation}\label{cov-single-prop}
\begin{aligned}
  {\cal E} &= \lim_{B\rightarrow A^c}\frac{3}{16G_{N}^{(d+1)}} \big[2{\cal Y}_{A}^{ext}+{\cal Y}_{B_1}^{ext}+{\cal Y}_{B_2}^{ext}-{\cal Y}^{ext}_{A\cup B_1}-{\cal Y}^{ext}_{A\cup B_2}\big],
  \end{aligned}
\end{equation}
where ${\cal Y}_{\gamma}^{ext}$  denotes the area of the extremal surface homologous to the corresponding subsystem $\gamma$ and $B=B_1\cup B_2 \to A^c$ is the bipartite limit such that the intervals $B_1$ and $B_2$ are extended to infinity and $B$ becomes $A^c$.

\subsection{Extremal rotating BTZ black hole}
Having described the computation of the covariant holographic entanglement negativity for various mixed states in a $CFT_{1+1}$ dual to bulk non-extremal BTZ black hole we now turn our attention to $CFT_{1+1}$s dual to bulk extremal rotating BTZ black holes. The metric for the bulk extremal rotating BTZ black hole is given as
\begin{equation}\label{ext-btz}
 ds^{2}=-\frac{(r^2-r_{0}^2)^2}{r^2}dt^2 + \frac{r^2 dr^2}{(r^2-r_{0}^2)^2}+ r^2(d\phi-\frac{r_{0}^2}{r^2}dt)^2.
\end{equation}
In the extremal limit $(r_+=r_-=r_0)$, the mass $M$ is equal to the angular momentum $J$ $(M=J=2 r_0^2)$ and the corresponding Hawking temperature vanishes ($T_H=0$). 

To compute the extremal entanglement wedge cross section in this geometry, we implement the following coordinate transformations which maps the extremal BTZ metric in eq. (\ref{ext-btz}) to the $\mathrm{AdS_3}$ metric in the Poincar\'{e} coordinates eq. (\ref{poincare-coord}) as
\begin{equation}\label{extbtzmaps}
\begin{aligned}
&w_{+}=\phi+t-\frac{r_0}{r^2-r_0^2}, ~~~w_{-}=\frac{1}{2r_0}e^{2r_0(\phi-t)},\\
& ~~~~~~~~~~~~~~Z=\frac{1}{\sqrt{r^2-r_0^2}}e^{r_0(\phi-t)}.
\end{aligned}
\end{equation}
The infra red cut-offs  $r_\infty$ and $\epsilon_i$ for the bulk in the BTZ and the Poincar\'{e} coordinates are related in this case by
\begin{equation}\label{cut-off-extremal}
\begin{aligned}
\varepsilon_{i}=\frac{1}{r_{\infty}}e^{r_0(\phi_{i}-t_0)}.
\end{aligned}
\end{equation}
\subsubsection{Covariant holographic negativity for adjacent intervals}
The extremal EWCS for the mixed state configuration of  
adjacent intervals $A$ and $B$ of lengths $l_1$ and $l_2$ may be computed in a similar manner as in subsection (\ref{subs-adj}) and may be expressed as
\begin{equation}
E_W^{ext}=\frac{c}{12}\ln\Bigg[\frac{l_1 l_2}{\epsilon(l_1+l_2)}\Bigg]
+\frac{c}{12}\ln\Bigg[\bigg(\frac{1}{r_0 \epsilon}\bigg)\frac{\sinh{\big(r_0l_1\big)}\sinh{\big(r_0l_2\big)} }{\sinh\{r_0(l_1+l_2)\}}\Bigg]+\frac{c}{6}\ln 4.
\end{equation}
It is now straightforward to obtain the corresponding covariant holographic entanglement negativity for the mixed state in question utilizing the proposal described in (\ref{cov-neg-proposal}) as follows
\begin{equation}\label{en-adj-ext}
\mathcal{E}=\frac{c}{8}\ln\Bigg[\frac{l_1 l_2}{\epsilon(l_1+l_2)}\Bigg]
+\frac{c}{8}\ln\Bigg[\bigg(\frac{1}{r_0 \epsilon}\bigg)\frac{\sinh{\big(r_0l_1\big)}\sinh{\big(r_0l_2\big)} }{\sinh\{r_0(l_1+l_2)\}}\Bigg]+\frac{c}{4}\ln 4.
\end{equation}
The above result also matches with the corresponding $CFT_{1+1}$ replica technique result in \cite{Jain:2017uhe} with the extra constant factor related to holographic Markov gap \cite{Hayden:2021gno}. We observe that the first term in the above expression is half of the ground state entanglement negativity for the mixed state of adjacent intervals whereas the second term resembles a finite temperature contribution to the entanglement negativity with an effective Frolov-Throne temperature $T_{\mathrm{FT}}=\frac{r_0}{\pi}$ \cite{Frolov:1989jh,Caputa:2013lfa}. 
As earlier the covariant holographic entanglement negativity in eq. (\ref{en-adj-ext}) may also be computed upto the constant factor from an earlier holographic proposal based on the algebraic sum of the lengths of bulk extremal curves homologous to the intervals and their appropriate combinations as described in eq. (\ref{CO-HEN-CONJ}) and
\cite{Jain:2017uhe}.

\subsubsection{Covariant holographic negativity for disjoint intervals}
For the case of two disjoint intervals, the extremal EWCS for the bulk extremal black hole geometry may once again be obtained in a manner similar to that described in subsection (\ref{subs-disj})
as
\begin{equation}
\begin{aligned}
E_W^{ext}(A:B) & = \frac{ c }{ 12 }
\ln \left [ \frac{
\left ( l_1 + l_s \right )
\left ( l_2 + l_s \right )
}{
l_s \left ( l_1 + l_2 + l_s \right )
} \right ] \\
& \quad + \frac{ c }{ 12 }
\ln \left [ \frac
{
\sinh \left \{ r_0 \left ( l_1 + l_s \right ) \right \}
\sinh \left \{ r_0 \left ( l_2 + l_s \right ) \right \}
}{
\sinh \left ( r_0 l_s \right )
\sinh \left \{ r_0 \left ( l_1 + l_2 + l_s \right ) \right \}
}
\right ] +\frac{c}{6} \ln 4.
\end{aligned}
\end{equation}
As earlier using the eq. (\ref{cov-neg-proposal}), the covariant holographic entanglement negativity for the mixed state of two disjoint intervals in proximity in the $CFT_{1+1}$ dual to a bulk extremal BTZ black hole may be expressed as
\begin{equation}\label{en-disj-ext}
\begin{aligned}
\mathcal{E} & = \frac{ c }{ 8 }
\ln \left [ \frac{
\left ( l_1 + l_s \right )
\left ( l_2 + l_s \right )
}{
l_s \left ( l_1 + l_2 + l_s \right )
} \right ] \\
& \quad + \frac{ c }{ 8 }
\ln \left [ \frac
{
\sinh \left \{ r_0 \left ( l_1 + l_s \right ) \right \}
\sinh \left \{ r_0 \left ( l_2 + l_s \right ) \right \}
}{
\sinh \left ( r_0 l_s \right )
\sinh \left \{ r_0 \left ( l_1 + l_2 + l_s \right ) \right \}
}
\right ] +\frac{c}{4} \ln 4.
\end{aligned}
\end{equation}
We note that modulo the constant term the above result matches exactly with the corresponding $CFT_{1+1}$ replica result \cite{Malvimat:2018ood}. Excluding the constant term in the above expression the first term is exactly half of the corresponding holographic entanglement negativity for the mixed state at zero temperature whereas the second term as earlier resembles a finite temperature correction at an effective Frolov-Throne temperature $ T_{ FT } = \frac{ r_0 }{ \pi } $ \cite{Frolov:1989jh,Caputa:2013lfa}. 
Once again the above result in eq. (\ref{en-disj-ext}) may also be obtained upto the constant factor from an alternative holographic proposal involving the algebraic sum of the lengths of bulk extremal curves homologous to appropriate combinations of the intervals as described in eq. (\ref{cohendisj}).

\subsubsection{Covariant holographic negativity for a single interval}

For the zero temperature mixed state of a single interval $A$ of length $l$ in the $CFT_{1+1}$ dual to the bulk extremal rotating BTZ black hole we proceed as in subsection (\ref{subs-single}) and obtain the corresponding extremal EWCS as follows
\begin{equation}
E_W^{ext}(A:B)=\frac{c}{6}\ln\left[\frac{l}{\epsilon}\right]+\frac{c}{6}\ln\left[\frac{\beta_{FT} }{\pi \epsilon}\sinh\left(\frac{\pi l}{\beta_{FT}}\right)\right]-\frac{\pi c l}{6 \beta_{FT}}+\frac{c}{3} \ln 4,
\end{equation}
where $\beta_{FT}$ is the Frolov-Thorne temperature given by $T_{FT}=\frac{1}{\beta_{-}}=\frac{r_0}{\pi}$.
We now employ the covariant proposal described in eq. (\ref{cov-neg-proposal}) to compute the holographic entanglement negativity for the mixed state in question as follows
\begin{equation}\label{en-ext-single}
{\cal E}=\frac{c}{4}\ln\left[\frac{l}{\epsilon}\right]+\frac{c}{4}\ln\left[\frac{\beta_{FT} }{\pi \epsilon}\sinh\left(\frac{\pi l}{\beta_{FT}}\right)\right]-\frac{\pi c l}{4 \beta_{FT}}+\frac{c}{2} \ln 4.
\end{equation}
Once again the above result excluding the constant term matches with the corresponding dual $CFT_{1+1}$ replica result \cite{Chaturvedi:2016opa} and may be expressed as
\begin{equation}
 {\cal E}=\frac{3}{2}\big[S_{A}-S^{FT}_{A}\big]+const.
\end{equation}
Note that in this case the covariant holographic entanglement negativity involves the removal of 
$S^{FT}_{A}=\frac{\pi c l}{6\beta_{-}}$ which is an effective thermal entropy of the subsystem $A$ at the Frolov-Thorne temperature $T_{FT}$ and corresponds to the entropy of degeneracy of the ground state. In the bulk this describes an emergent thermodynamic behaviour for the rotating extremal BTZ black hole at zero temperature with a non zero entropy of degeneracy. This indicates that the entanglement negativity which should lead to an upper bound on the distillable entanglement treats the entropy of degeneracy as an effective thermal contribution which is subtracted out. Once again this is a universal result and may also be computed upto the constant from an earlier alternative holographic proposal involving bulk extremal curves homologous to appropriate combinations of intervals as described  in eq. (\ref{cov-single-prop})\cite{Chaturvedi:2016opa}.

\section{Time dependent holographic entanglement negativity}\label{sec6}

In the earlier section we computed the covariant holographic entanglement negativity for various bipartite mixed states in a $CFT_{1+1}$ dual to a bulk stationary configuration described by a rotating BTZ black hole. In this section we turn our attention to the covariant holographic entanglement negativity for time dependent mixed states in $CFT_{1+1}$ dual to time dependent bulk $\mathrm{AdS_3}$ geometry described by a Vaidya-$\mathrm{AdS_3}$ configuration.
This bulk $\mathrm{AdS_3}$ geometry is one of the simplest examples of a time-dependent configuration which describes an idealized gravitational collapse and subsequent black hole formation \cite {Anous:2016kss,Anous:2017tza}.
The metric for the Vaidya-$\mathrm{AdS_3}$ in the Poincar\'e coordinates is given by \cite{Hubeny:2007xt,AbajoArrastia:2010yt}
\begin{equation}
ds^2=-\left[r^2 -m(v)\right]~dv^2 +2~dr~dv +r^2 ~dx^2,
\end{equation}
where $m(v)$ is a function of the (light cone) time $v$. For a constant mass function $m(v)$, the above metric reduces to a non-rotating BTZ black hole described in eq. (\ref{static-btz}).

To obtain the extremal entanglement wedge cross section for this bulk geometry, we use the adiabatic approximation in which the mass function $m(v)$ is
assumed to be a slowly varying function of $v$ and $\dot{m}(v)\ll 1$ as described in \cite{Hubeny:2007xt}. In this approximation we first compute the extremal entanglement wedge cross section in the static BTZ black hole geometry and then treat the mass as a function of the (light cone) time $v$. In what follows we utilize our covariant holographic proposal to compute the holographic entanglement negativity for various mixed state configurations in a $CFT_{1+1}$ which is dual to this time dependent bulk geometry. Note that we continue to assume that the backreaction parameter $\mathcal{X}_2 =\frac{3}{2}$ remains unchanged for the $AdS_3/CFT_2$ scenario as the bulk extremal surfaces are extremal curves (geodesics) for both static and non static bulk geometries.

\subsection{Entanglement negativity for adjacent intervals}

For the mixed state of adjacent intervals in a $CFT_{1+1}$ dual to the bulk static BTZ black hole the minimal entanglement wedge cross section was described in eq. (\ref{ew-adj-finite}). We now use the adiabatic approximation where mass $m$ can be regarded as a time dependent function $m(v)$ and $\beta=2\pi/\sqrt{m}$ in eq. (\ref{ew-adj-finite}) to obtain the extremal EWCS for the bulk Vaidya-$\mathrm{AdS_3}$ geometry as follows
\begin{equation}\label{EW-vaidya-adj}
\begin{aligned}
E_W^{ext}(A:B)=\frac{1}{8 G_N^{(3)}}
\ln\bigg[4 r_\infty^2\frac{\sinh^2\left(\sqrt{m(v)}~l_1/2~\right)~
\sinh^2\left(\sqrt{m(v)}~l_2/2~\right)}{\sinh^2\left(\sqrt{m(v)}~(l_1+l_2)/2~\right)m(v)}\bigg]+\frac{1}{4 G_N^{(3)}}\ln 4,
\end{aligned}
\end{equation}
where the  UV cut-off is $r_{\infty}=1/\varepsilon$. Now the covariant holographic entanglement negativity for the mixed state of adjacent intervals may be computed by using the holographic construction from eq. (\ref{cov-neg-proposal}) and is expressed as
\begin{equation}\label{en-vaidya-adj}
\begin{aligned}
\mathcal{E}=\frac{3}{ 8 G_N^{(3)}} \ln [2 ~r_{\infty}]+
\frac{3}{ 16 G_N^{(3)}}\ln\bigg[\frac{\sinh^2\left(\sqrt{m(v)}~l_1/2
~\right)~\sinh^2\left(\sqrt{m(v)}~l_2/2~\right)}{\sinh^2\left(\sqrt{m(v)}~(l_1+l_2)/2~\right)m(v)}\bigg]+\frac{3}{8 G_N^{(3)}}\ln 4.
\end{aligned}
\end{equation}
We observe as earlier that the above result matches with the corresponding result modulo the constant obtained from an earlier alternate holographic construction \cite{Jain:2017uhe} involving bulk extremal curves described in eq. (\ref{CO-HEN-CONJ}) which serves as a consistency check. Note however that the finite part 
$\mathcal{E}_{fin}$ in the above eq. (\ref{en-vaidya-adj}) obtained by subtracting the divergent part and the constant term becomes negative when plotted against $v$ contradicting the quantum information theory expectation of 
positive semi definite entanglement measures.  As described in \cite{Jain:2017uhe} this is an artifact of the cut-off regularization and may be resolved by computing the renormalized entanglement negativity 
$\mathcal{E}_{ren}=l\frac{\partial \mathcal{E}}{\partial l}$ which leads to a non-negative and a
monotonically decreasing function of $v$ that saturates to a fixed small value for large $v$. This indicates the thermalization of the mixed state which corresponds to the bulk black hole formation and is also consistent with quantum information theory expectations.

\subsection{Entanglement negativity for disjoint intervals}
The extremal EWCS for the mixed state configuration of two disjoint intervals
in proximity in the dual $CFT_{1+1}$ may be obtained from eq. (\ref{ew-disj-zero}) by employing the conformal transformation $z \to w=(\beta/2\pi)\ln z$ from the plane to the cylinder followed by the
adiabatic approximation and using $\beta=2\pi/\sqrt{m}$ as follows
\begin{equation}
\begin{aligned}
 E_W^{ext}(A:B) = \frac{ 1 }{ 8 G_N^{(3)} }
 \ln
\left [ \frac
{
\sinh^2 \left \{ \sqrt{ m(v)} \left ( l_1 + l_s \right ) / 2  \right \}
\sinh^2 \left \{ \sqrt{ m(v)} \left ( l_2 + l_s \right ) / 2  \right \}
}{
\sinh^2 \left ( \sqrt{ m(v)} l_s/2  \right )
\sinh^2 \left \{ \sqrt{ m(v)} \left ( l_1 + l_2 + l_s \right ) / 2  \right \}
}
\right ] +\frac{1}{4 G_N^{(3)}}\ln 4.
\end{aligned}
\end{equation}
Utilizing our proposal described by eq. (\ref{cov-neg-proposal}) and the above expression, the covariant holographic entanglement negativity for  the mixed state in  question may be computed as follows
\begin{equation}
\begin{aligned}
\mathcal{E} = \frac{ 3 }{ 16 G_N^{(3)} }
 \ln
\left [ \frac
{
\sinh^2 \left \{ \sqrt{ m(v)} \left ( l_1 + l_s \right ) / 2  \right \}
\sinh^2 \left \{ \sqrt{ m(v)} \left ( l_2 + l_s \right ) / 2  \right \}
}{
\sinh^2 \left ( \sqrt{ m(v)} l_s/2  \right )
\sinh^2 \left \{ \sqrt{ m(v)} \left ( l_1 + l_2 + l_s \right ) / 2  \right \}
}
\right ] +\frac{3}{8 G_N^{(3)}}\ln 4.
\end{aligned}
\end{equation}
Note that the above result is cut off independent. Since the above expression has no divergent part, as shown in \cite{Malvimat:2018ood} when plotted against $v$ the entanglement negativity decreases monotonically and saturates to a small value for large $v$ characterizing the thermalization of the mixed state in question and corresponds to the black hole formation in the bulk $\mathrm{AdS_3}$ geometry. We also observe that excluding the constant term, the above result matches with the corresponding result determined from an alternative holographic proposal  based on the algebraic sum of the lengths of appropriate bulk extremal curves described in \cite{Malvimat:2018ood}.

\subsection{Entanglement negativity for a single interval}
In an exactly similar fashion as in the previous subsection, we utilize the adiabatic approximation in eq. (\ref{ew-BTZ-single}) to compute the bulk extremal EWCS for a single interval $A$ of length $l$ in the dual $CFT_{1+1}$ as
\begin{equation}
E_W^{ext}(A:B)=\frac{1}{4G_N^{(3)}}\left(\ln\left[\frac{4 r_\infty^2}{m(v)}\sinh^2\left(\frac{\sqrt{m(v)}~l}{2}\right)\right]-\sqrt{m(v)}~l\right)+\frac{1}{2 G_N^{(3)}}\ln 4,
\end{equation}
where the UV cut-off  $r_{\infty}=1/\varepsilon$.
We now employ our covariant construction described in eq. (\ref{cov-neg-proposal}) to obtain the covariant holographic entanglement negativity for the mixed state of a single interval in a dual $CFT_{1+1}$ as follows
\begin{equation}
\mathcal{E}=\frac{3}{4G_N^{(3)}}\ln(2 r_\infty)+
\frac{3}{8G_N^{(3)}}\left(\ln\left[\frac{\sinh^2(\sqrt{m(v)}~l/2)}{m(v)}\right]-\sqrt{m(v)}~l\right)+\frac{3}{4 G_N^{(3)}}\ln 4.
\end{equation}
As is now evident the above result modulo the constant matches with the covariant entanglement negativity of a single interval obtained from the alternate holographic proposal  described in eq. (\ref{cov-single-prop}) \cite{Chaturvedi:2016opa}. Similar to the earlier case, the renormalized entanglement negativity is non-negative, and decreases monotonically with $v$ saturating to a small fixed value at large $v$ which characterizes the thermalization for the mixed state and describes the formation of a black hole in the bulk.

\section{Summary and discussion}\label{sec7}
To summarize, we have proposed a covariant holographic entanglement negativity construction for bipartite states in $CFT_{1+1}$s dual to non-static bulk $AdS_3$ geometries which is based on the extremal entanglement wedge cross section (EWCS). Our proposal has been substantiated by utilizing our construction to compute the covariant holographic entanglement negativity for bipartite mixed states in $CFT_{1+1}$s dual to bulk stationary geometries described by rotating non extremal and extremal BTZ black holes. The results for the covariant holographic entanglement negativity for these mixed states in the dual $CFT_{1+1}$ obtained from our construction matches with the corresponding replica technique results modulo certain constants related to the Markov gap in the large central charge limit. This constitutes 
robust consistency checks for our covariant holographic construction based on the extremal EWCS for the $AdS_3/CFT_2$ scenario. As a further substantiation we have subsequently employed our covariant construction to compute the holographic entanglement negativity for bipartite mixed states in $CFT_{1+1}$s dual to bulk time dependent configuration described by the Vaidya-$AdS_3$ geometries. It is observed that our results for the time dependent scenarios
matched modulo certain constants related to the Markov gap with the corresponding results obtained through an earlier holographic construction involving the algebraic sum of the lengths of bulk extremal curves homologous to certain combinations of the intervals. These provide additional consistency checks of our construction for the $AdS_3/CFT_2$ scenario.

It is significant to note that the covariant holographic entanglement negativity obtained from our proposal, for  bipartite mixed states in $CFT_{1+1}$s dual to bulk stationary and time dependent configurations involving the extremal entanglement wedge cross section match upto certain constant factors with those obtained from an earlier holographic construction involving algebraic sums of the lengths of bulk extremal curves homologous to certain appropriate combination of the intervals appropriate to the bipartite state in question. Exactly similar conclusions have been described in \cite {Basak:2020oaf} for the holographic entanglement negativity for bipartite mixed states in $CFT_{1+1}$s dual to static bulk $AdS_3$ geometries obtained from the minimal EWCS. This clearly demonstrates the equivalence
of the two holographic constructions for the $AdS_3/CFT_2$ scenario modulo the constants related to the Markov gap mentioned above. It is an interesting open issue to extend this construction based on the EWCS to generic higher dimensional $AdS_{d+1}/CFT_d$ scenarios. In this context note that the computation of the entanglement wedge 
in higher dimensions is extremely non trivial and the alternate holographic proposal based on the
algebraic sum of the areas of bulk extremal surfaces homologous to appropriate combination of the subsystems may provide a simpler approach. We should also mention here that an explicit analytic continuation to determine the backreaction of the cosmic brane arising from the bulk conical defect on the extremal EWCS is still a non trivial open issue for the covariant $AdS_{d+1}/CFT_d$ scenarios. An argument for this may be developed following the maximin construction that demonstrates the equivalence of the RT and the HRT construction. This is an extremely interesting problem for future investigations.

We emphasize here that our covariant holographic construction based on the extremal entanglement wedge cross section provides an elegant extension of  the proposal described in \cite{Kudler-Flam:2018qjo, Kusuki:2019zsp} to time dependent mixed state configurations in  $CFT_{1+1}$s dual to non static bulk geometries in the $AdS_3/CFT_2$ framework.  The entanglement wedge cross section has also been shown to be related to other quantum information measures such as the entanglement of purification, odd entanglement entropy and the reflected entropy. It would be extremely interesting to explore the relation between these entanglement measures especially the reflected entropy and the entanglement wedge cross section for bipartite mixed states dual to non static bulk geometries considered in this work. Our covariant holographic entanglement negativity construction is also expected to find applications to the study of time dependent entanglement issues in quantum gravity and many body condensed matter physics in the context of the $AdS_3/CFT_2$ scenario. These serve as fascinating open issues for future investigations.

\section{Acknowledgement}
We would like to thank Vinay Malvimat and Debarshi Basu for useful discussions and suggestions.

\bibliographystyle{utphys}

\bibliography{references}

%\bibliography{references}

\end{document}